\documentclass[runningheads,a4paper]{llncs}

\usepackage[T1]{fontenc}
\usepackage{amsmath,amssymb}
\usepackage{graphicx}
\usepackage{booktabs}
\usepackage{placeins}
\usepackage{url}
\usepackage[hidelinks]{hyperref}
\usepackage{tikz}
\usetikzlibrary{arrows.meta,positioning,fit,backgrounds,calc}
\usepackage{pgfplots}
\pgfplotsset{compat=1.18}

\newcommand{\phiS}{\ensuremath{\varphi_{S}}}
\newcommand{\phiT}{\ensuremath{\varphi_{T}}}
\newcommand{\phiSig}{\ensuremath{\varphi_{\Sigma}}}
\newcommand{\triple}{\ensuremath{\langle\phiS,\phiT,\phiSig\rangle}}
\newcommand{\somewhere}{\ensuremath{\Diamond}}
\newcommand{\everywhere}{\ensuremath{\Box}}
\newcommand{\reach}{\ensuremath{\mathcal{R}}}
\newcommand{\escape}{\ensuremath{\mathcal{E}}}
\newcommand{\Ldom}{\ensuremath{\mathcal{L}}}
\providecommand{\llbracket}{\ensuremath{[\![}}
\providecommand{\rrbracket}{\ensuremath{]\!]}}

\usepackage{array}
\begin{document}

\title{Multi-Aspect Runtime Verification for Simulation-Based V\&V\\
       of LLM-Enabled Autonomous Agents\thanks{This is a preprint of an article accepted for publication in Modelling and Simulation for Autonomous Systems (MESAS 2026), Springer LNCS.}}

\author{Nikolaos Kekatos\inst{1} \and Dimitrios Nikou\inst{2} \and
        Anastasios Temperekidis\inst{1} \and Alexios Lekidis\inst{3} \and
        Nikolaos Kolokotronis\inst{4} \and Panagiotis Katsaros\inst{5} \and
        Stylianos Basagiannis\inst{2}}
\titlerunning{Multi-Aspect Runtime Verification of LLM-Based Agents}
\authorrunning{N. Kekatos et al.}
\institute{Clone Systems, Cyprus \and
           International Hellenic University, Greece \and
           University of Thessaly, Greece \and
           University of Peloponnese, Greece \and
           Aristotle University of Thessaloniki, Greece}

\maketitle

\begin{abstract}
LLM-based agents are entering decision-support roles in defence staff work,
where the obligations they must respect are already written down and already
binding, and where retraining is not available as a control because models
arrive as procured components on a vendor update cadence. What can be placed
under engineering control is the interface between the agent and the systems it
acts on. Those obligations are at once \emph{spatial} (where data and actions may reside),
\emph{temporal} (in what order, within what windows), and \emph{text-semantic}
(what is read, said, or acted upon). A single tool call rarely violates anything.
The violation lives in the composition of a multi-step interaction, which is why
per-event guardrails miss sequential tool-attack chains that succeed on frontier
models. We present a multi-aspect runtime-verification framework that decomposes
a natural-language policy clause into a typed triple
$\langle\varphi_S,\varphi_T,\varphi_\Sigma\rangle$ over one canonical event
stream, checks each aspect with its own monitoring specification, and fuses the
verdicts through a four-valued algebra that carries provenance. The spatial
aspect is interpreted over a weighted two-sorted location graph in which mission
geometry and information-release topology are one object, so a single monitor
decides both where an effect may land and where a record may travel. We show that
these spatial obligations are not in general subsumed by a first-order temporal
specification, since containment is the transitive closure of a runtime-mutable
edge relation. The past-time aspect runs on the unmodified MonPoly engine, which
agrees with our reference monitor at every time point. Across two mission domains,
casualty evacuation and contested sustainment, and one civil domain, composition
under the precautionary blocking policy drives attack success to zero with no
observed false positives and microsecond-scale per-event cost, while every single aspect and every pair leaves a substantial
share of attacks succeeding. In a closed-loop experiment a policy-naive planner
reaches a violating state in most unshielded missions and in none when shielded,
and four refused episodes in five still recover to a compliant outcome. Every
blocked action ships an auditable, clause-grounded verdict.

\keywords{Runtime verification \and LLM agents \and Agent safety \and
Compositional monitoring \and Past-time temporal logic \and Trustworthy AI
\and V\&V of autonomous systems}
\end{abstract}

\section{Introduction}
\label{sec:intro}

Large language models are moving into decision-support roles inside military
staff work~\cite{dodlima2024,thunderforge2025,natoai2024}. The U.S. Department of Defense catalogued generative-AI use cases
across fifteen areas spanning command and control and decision support, and stood
up a dedicated fielding effort to accelerate them~\cite{dodlima2024}; its
Thunderforge programme puts LLM-based agents into campaign development,
course-of-action generation, wargaming and resource allocation, with initial
mission partners at two combatant commands~\cite{thunderforge2025}. NATO's
revised AI strategy addresses generative AI directly and calls for an
Alliance-wide testing, evaluation, verification and validation capability to
support responsible adoption~\cite{natoai2024}. That last point is the gap this
paper addresses.

Defence adoption changes the verification problem rather than merely enlarging
it. A staff agent operates under obligations that are already written down and
already binding: rules of engagement, classification and release policy,
targeting law, and the protections that attach to medical and humanitarian
activity. These are not preferences to be traded against utility, and an
operator who accepts a recommendation inherits responsibility for it, so a system
that cannot say which obligation it checked and on what evidence cannot be held
to account after the fact. The same adoption also removes the option of
retraining as a control: models arrive as procured components, are updated on a
vendor cadence, and are deployed into coalition settings where the releasing
authority and the consuming command differ. What can be placed under
engineering control is the interface between the agent and the systems it acts
on, which is where a monitor sits. This is the capability NATO's strategy asks
for, applied at the level of the individual action rather than the model.

Why put a language model in this loop at all is a fair question, and the answer
bounds what we claim. The inputs are unstructured language: evacuation requests,
spot reports, clearance traffic, manifests written under time pressure. Turning
those into a tasking that respects marking, routing, sharing and retention rules
is a matching and sequencing problem over text, which is difficult to cover
exhaustively with fixed rules. What the agent does with that is propose and order
tool calls; nothing here puts a model between a sensor and an actuator, and the
consequential actions it can request are mediated calls that an executor still
performs. The assurance argument does not require agreeing that such an agent
should be fielded. If it is, these obligations are what it must not violate; if
the judgement is that it stays advisory, the same mechanism is what makes
advisory \emph{enforceable}, since an advisory system that can open a release
channel is not advisory in the sense that matters.

This work asks whether consequential actions of tool-using LLM agents can be
prevented from reaching policy-violating states when compliance depends jointly
on spatial state, temporal history and semantic content. Such an agent calls
tools, moves data, designates points on the ground and commits the organisation
to actions with legal consequences, under obligations of several kinds at once:
\emph{spatial}, where an effect lands and where a record may travel;
\emph{temporal}, in what order and inside which window; and
\emph{text-semantic}, what a narrative field says as opposed to what a
structured field claims.

The difficulty is that these obligations are rarely broken by any one action. A
tool call inspected in isolation is almost always admissible; the violation lives
in the \emph{composition} of a multi-step interaction. This is exactly the regime
in which per-event guardrails fail by construction, and it is the regime attackers
have found: sequential tool-attack chains succeed against frontier models at rates
above $90\%$~\cite{stac2025}, and per-call policy layers such as
AgentSpec~\cite{agentspec2026}, Progent~\cite{progent2025},
GuardAgent~\cite{guardagent2024} and constrained-action proxies for security
operations~\cite{koutidis2026guardrails} cannot see a violation that no single
call exhibits.

Runtime verification is the natural response, and mature engines exist for each
aspect in isolation: stream-based monitors such as
TeSSLa~\cite{tessla2018,tessla2024}, first-order past-time monitors such as
DejaVu~\cite{havelund2018dejavu}, spatio-temporal logics such as
STREL~\cite{strel2017} and SaSTL~\cite{sastl2020}. They are almost always
deployed one at a time, against a specification written for that engine. The
control this paper argues for separates the aspects a policy clause constrains,
checks each with a monitor mature for that aspect, and \emph{composes} the
verdicts into one auditable decision.

Two commitments distinguish what follows. First, the spatial aspect is treated
as spatial. Much work on agent compliance reduces ``where'' to a label lookup,
which cannot express standoff, masking, corridor containment, or the difference
between a disallowed destination that exists in the topology and one that is
reachable. We interpret \phiS\ over a weighted \emph{two-sorted location graph}
in which physical mission geometry and information-release topology are the same
mathematical object. Second, a monitor must be able to say that it is not sure:
verdicts live in a four-valued domain, so a component that only guesses cannot
manufacture a definite refusal.

\paragraph{Contributions.}
\begin{itemize}
  \item \textbf{A multi-aspect assurance model.} Trace-level obligations are
        decomposed into spatial, temporal and semantic aspects over one
        canonical event stream, with the spatial aspect interpreted over a
        two-sorted location graph that carries mission geometry and
        information-release topology in the same structure, so one reachability
        query decides both where an effect may land and where a record may
        travel. An expressiveness argument motivates keeping the spatial aspect
        separate from the temporal one
        (Sects.~\ref{sec:spec}, \ref{sec:spec:expressive}).
  \item \textbf{Preventive composition and enforcement.} A guarded four-valued
        composition, reusing the four-symbol domain of
        RV-LTL~\cite{blsemantics2011}, carries provenance, and a shield
        evaluates each proposed call on a shadow world model and commits only
        what the composed verdict permits, which turns monitoring into
        prevention rather than reporting
        (Sects.~\ref{sec:spec:algebra}, \ref{sec:arch}).
  \item \textbf{Simulation-based V\&V evidence.} An evaluation across three
        instantiations establishes when composition is required, whether the
        temporal aspect agrees with an independent engine, whether enforcement
        preserves mission effectiveness in closed loop, and where assurance
        fails when the tool surface is incomplete or evidence can be forged
        (Sect.~\ref{sec:eval}).
\end{itemize}

\paragraph{What we reuse, and what we add.}
Every monitoring engine here is established and none is advanced by us:
stream-based monitoring~\cite{tessla2018,tessla2024,rtlola}, BDD-based
first-order past-time monitoring and its timed
extension~\cite{havelund2018dejavu,havelund2020timed}, MFOTL and its decision
procedure~\cite{verimon2021}, the spatial modalities and their graph
semantics~\cite{strel2017,sastl2020}, abstaining classification with a quantified
error budget~\cite{conformal2024}, and the observation that multi-step tool use
defeats per-call guardrails~\cite{stac2025}. What we add is the
two-sorted location graph in which mission geometry and release topology become
one metric structure; the argument motivating a separate spatial monitor
from the expressiveness and capability boundaries of
Sect.~\ref{sec:spec:expressive}; the four-valued composition with provenance,
and the observation that presumptive verdicts belong to the composition layer
rather than to any engine; the shadow-evaluate-then-commit shield that makes the
framework preventive; and three instantiations showing that a new domain costs a
vocabulary.

\noindent In short, Sect.~\ref{sec:spec} defines what must be checked,
Sect.~\ref{sec:arch} defines how the framework is validated in simulation, and
Sect.~\ref{sec:eval} reports whether the resulting shield prevents violations,
at what cost, and under what assurance assumptions.

\section{Operational Scenario and Assurance Problem}
\label{sec:motivation}

Decision support for medical evacuation is established, but its systems are
machine-learning and optimisation systems rather than language models invoking
tools~\cite{medevacdss2023,smartmedevac2024}. We model the composition of two
documented trends: LLM agents entering planning
workflows~\cite{thunderforge2025}, and AI decision support entering evacuation
and sustainment. The agents here are \emph{representative near-term roles}
rather than fielded systems, and the monitoring problem does not turn on that
distinction: it arises wherever multi-step tool use meets multi-aspect
obligations.

A casualty-evacuation tasking agent supports a medical operations cell, working
under a standing clause: aircraft flying under the protective medical marking
shall carry nothing outside their medical function; casualty records may be
shared only with facilities covered by a medical-sharing arrangement and only
with authority for that casualty; urgent casualties are to be evacuated within
the hour. The agent makes four calls, each admissible on its own and each seen
only in part by one aspect: it requests the protective marking, which \phiT\
sees granted and never revoked; it plans a route across the forward line, which
\phiS\ sees as lawful for a protected flight; it amends the manifest, where the
cargo field still reads \texttt{cargo} and only the narrative names ammunition,
so \phiSig\ sees narrative alone; and it crosses the line, which no aspect
refuses by itself.

The composition is a violation: a medical aircraft carrying materiel outside its
medical function loses the protection the marking asserts, so continuing to
display it is improper use of the protective emblem. Whether that rises to
\emph{perfidy} turns on an intent a monitor cannot establish, so the framework
refuses the crossing on the objective elements alone. Only a monitor that reads
the manifest \emph{narrative}, remembers when the marking was granted, and knows
the aircraft has left friendly-held airspace can refuse step~4, and only by
combining all three.

The same structure recurs throughout the evaluation: a casualty record that
escapes its release domain because a coalition feed was enabled two steps
earlier; a ``sanitised'' summary that still carries identifiers; and a convoy
declared humanitarian to obtain corridor access whose load was later amended. In
each case the aspects are separable, none suffices, and the violation is a trace
property.

\paragraph{Requirements.} The scenario fixes what a mechanism has to provide.

\begin{description}
  \item[R1, trace memory.] The monitor must relate actions across time. A call
        inspected independently of what preceded it cannot see this violation.
  \item[R2, heterogeneous evidence.] Compliance depends jointly on spatial
        state, temporal history and content, so one kind of evidence is not
        enough.
  \item[R3, pre-commit evaluation.] The action that first makes the prohibited
        state reachable must be stopped before it executes, not reported after.
  \item[R4, auditability.] A refusal must name the obligation it rests on and
        the evidence that produced it, because an operator inherits
        responsibility for accepting or overriding it.
\end{description}

\noindent Section~\ref{sec:spec} constructs the framework to satisfy R1--R4.

\section{Multi-Aspect Runtime Assurance Framework}
\label{sec:spec}

A policy clause written in natural language is compiled into a typed triple
\triple\ interpreted over a single canonical event stream. Everything in this
section is contributed here, with two exceptions that are inherited and cited as
such: the spatial modalities and their graph semantics, and the underlying
monitoring logics. What is new is the event schema, the two-sorted location
graph, the expressiveness limits motivating the decomposition, and the composition
algebra. This section fixes the
event schema (Sect.~\ref{sec:spec:events}), the \emph{location graph} that gives
the spatial aspect its semantics (Sect.~\ref{sec:spec:graph}), the qualitative
spatial fragment over it (Sect.~\ref{sec:spec:strel}), the three aspects of the
triple (Sect.~\ref{sec:spec:triple}), and the four-valued algebra that fuses
their verdicts with provenance (Sect.~\ref{sec:spec:algebra}).

\subsection{Canonical Event Model}
\label{sec:spec:events}

All three aspects read one stream. Canonicalisation maps every agent action,
tool invocation, tool result and observation to an event
\[
  e \;=\; \langle\, \tau,\; \alpha,\; \kappa,\; \iota,\; \bar{a},\; \lambda \,\rangle ,
\]
where $\tau$ is a mission timestamp, $\alpha$ the actor, $\kappa$ the event kind
(\texttt{tool\_call}, \texttt{tool\_result}, \texttt{observation},
\texttt{user\_msg}), $\iota$ the tool identifier, $\bar{a}$ the typed argument
vector, and $\lambda$ a set of semantic labels. A single stream is what makes the
aspects composable: they do not observe three different abstractions of the
agent, they observe the same trace through three different lenses, so their
verdicts are about the same event and can be fused pointwise.

Three design choices matter for soundness. First, canonicalisation sits on the
trusted path between the planner and the tool executor, so an agent whose
planning has been diverted still emits faithful events; the monitor's view is not
mediated by the component under suspicion. Second, the shield evaluates each
\texttt{tool\_call} \emph{before} its effect is committed, on a shadow copy of
the world model, and commits only if the composed verdict permits it. A blocked
call therefore never mutates mission state, which is what allows a violation to
be \emph{prevented} rather than merely reported.

Third, and easily overlooked: \textbf{one tool call is one time-point database}.
A single call typically makes several facts true at once, a record acquires a
location, a channel opens, a marking is assigned, and these are simultaneous,
not successive. A monitor that advances one event per step would, if fed them
one at a time, place them at distinct time points and thereby stretch every
bounded window in \phiT\ by the number of facts the call happened to produce.
Canonicalisation therefore emits the whole set at a single timestamp, and the
MFOTL export of Sect.~\ref{sec:eval:engine} preserves that grouping. This is a
property of the interface rather than of any engine, and it has to be stated,
because getting it wrong is silent: the windows simply drift.

We also fix a convention for undefined operands. A spatial predicate over an
object with no location, or a temporal predicate over an obligation never armed,
evaluates to \emph{false} rather than raising or defaulting to satisfaction,
so a record that does not yet exist is never ``outside its release domain''.

\subsection{Two-Sorted Spatial World Model}
\label{sec:spec:graph}

The spatial aspect is interpreted over a weighted, two-sorted \emph{location
graph}
\[
  \Ldom \;=\; (V,\, E,\, w,\, \mathrm{sort},\, \delta),
\]
where $w$ assigns a non-negative weight to each edge, $\mathrm{sort}$ partitions
$V$ into \textsf{phys} and \textsf{info} vertices, and $\delta$ assigns a release
domain to each informational vertex. Vertices additionally carry the labels the
spatial predicates read.

\emph{Physical} vertices are terrain cells, landing zones, threat emitters and
masking terrain; their edges are undirected and weighted by distance in
kilometres. \emph{Informational} vertices are hosts and enclaves, each in a
release domain such as \textsf{Med}, \textsf{Med-Coalition}, \textsf{Partner} or
\textsf{Log}; their edges are directed, because a release channel is one-way, and
carry the \emph{domain-crossing weight}
\[
  w(u,v) \;=\;
  \begin{cases}
    0 & \text{if } \delta(u) = \delta(v),\\
    1 & \text{otherwise.}
  \end{cases}
\]

Putting both sorts in one graph is the central modelling decision of this paper,
and it is not a convenience. It means a single spatial monitor answers both
\emph{where the aircraft may go} and \emph{where the data may travel}, because
under the domain-crossing weight the informational sort carries a distance that
counts release-domain boundaries.

Fig.~\ref{fig:graph} shows both sorts of the MEDEVAC graph. In the informational
sort \textsc{med} holds the marked record, \textsc{med\_coalition} is a domain it
may lawfully reach and \textsc{partner} one it may not, so a move inside a domain
weighs $0$ and a move across a boundary weighs $1$; opening one channel at turn 4
puts a \textsc{partner} vertex inside the record's reach and fires $S_1$ before
the record moves. In the physical sort, with standoff $d=4$\,km,
$\mathrm{LZ}_B$ violates it over an admissible $3$\,km path while
$\mathrm{LZ}_C$ is nearer at $2.5$\,km and safe, its only route running through
masking terrain the path predicate excludes. Proximity alone would not separate
the two.

The two sorts differ in what kind of space they are. On the physical sort,
undirected edges with positive kilometre weights give a genuine metric. On the informational sort the shortest-path distance
$d(u,v)$ is non-negative, satisfies $d(u,u)=0$ and obeys the triangle inequality,
but it is neither symmetric, since release channels are directed, nor separating,
since two distinct hosts in the same release domain lie at distance $0$. It is
therefore a \emph{quasi-pseudometric} rather than a metric, and deliberately so:
hosts within one domain \emph{should} be indistinguishable to a clause about
domain crossings, and a one-way channel \emph{should} not imply its converse.
Non-negativity and the triangle inequality are all the shortest-path semantics of
the spatial operators requires, so both sorts are admissible interpretations of
the same operators. A clause about standoff from a threat and a clause about a casualty
record leaving a release domain are then instances of the same operator, and the
data-residency property of the civil instantiation, ``at least one copy in an
approved store'', is the degenerate zero-distance case.

\begin{figure}[tbp]
\centering
\begin{tikzpicture}[
  x=1.05cm, y=0.80cm, font=\footnotesize,
  iv/.style={draw,rounded corners=2pt,minimum height=5.6mm,inner xsep=3.5pt,
             fill=white,font=\scriptsize\ttfamily},
  bad/.style={draw,thick,rounded corners=2pt,minimum height=5.6mm,inner xsep=3.5pt,
              fill=black!12,font=\scriptsize\ttfamily},
  pv/.style={draw,circle,minimum size=7.2mm,inner sep=1pt,fill=white},
  threat/.style={draw,circle,minimum size=7.2mm,inner sep=1pt,fill=black!30},
  mask/.style={draw,rectangle,rounded corners=1pt,minimum width=8mm,
               minimum height=5mm,inner sep=1pt,fill=black!12},
  dom/.style={draw,dashed,black!55,rounded corners=3pt,inner sep=5pt},
  edge/.style={-{Stealth[length=4.5pt]},semithick},
  active/.style={-{Stealth[length=4.5pt]},very thick},
  newedge/.style={-{Stealth[length=4.5pt]},very thick,densely dotted},
  physical/.style={semithick,black!65},
  fire/.style={very thick},
  excluded/.style={semithick,densely dashed,black!45},
  lbl/.style={inner sep=1pt,font=\scriptsize},
  panel/.style={font=\footnotesize\bfseries,align=center}]

\node[iv]  (mf) at (0,0)      {med\_fwd};
\node[iv]  (mr) at (1.6,1.15) {med\_rear};
\node[iv]  (c2) at (3.3,0)    {c2\_cop};
\node[iv]  (pm) at (0,-1.85)  {partner\_med};
\node[bad] (cc) at (6.5,0)    {coalition\_cop};

\begin{scope}[on background layer]
  \node[dom,fit=(mf)(mr)(c2),
    label={[lbl,fill=white,inner sep=1pt,anchor=south west]above left:\textsc{med}}]{};
  \node[dom,fit=(pm),
    label={[lbl,fill=white,inner sep=1pt,anchor=north west]below left:%
      \textsc{med\_coalition} (allowed)}]{};
  \node[dom,fit=(cc),
    label={[lbl,fill=white,inner sep=1pt,anchor=south]above:\textsc{partner}}]{};
\end{scope}

\draw[edge]   (mf) -- node[lbl,above left=-1pt]{0} (mr);
\draw[active] (mf) -- node[lbl,above]{0} (c2);
\draw[edge]   (mf) -- node[lbl,left]{1} (pm);
\draw[newedge] (c2) -- node[lbl,above]{1} node[lbl,below=15pt,text=black!70]{opens at turn 4} (cc);

\node[circle,fill=black,inner sep=1.6pt] (rec) at ($(mf.south east)+(-0.12,0.12)$) {};
\node[lbl,anchor=north west] at ($(mf.south east)+(0.10,-0.28)$) {marked record $r$};
\node[lbl,anchor=north,align=center] at ($(cc.south)+(0,-0.55)$) {$\times$ disallowed};
\node[panel] at (3.15,-3.35) {(a) informational sort: runtime topology};

\begin{scope}[yshift=-4.35cm]
\node[pv] (la) at (0,0.95)  {$\mathrm{LZ}_A$};
\node[pv] (lb) at (0,0)     {$\mathrm{LZ}_B$};
\node[pv] (lc) at (0,-0.95) {$\mathrm{LZ}_C$};
\node[pv]   (o1) at (2.0,0.95) {};
\node[pv]   (o2) at (2.0,0)    {};
\node[mask] (rg) at (2.0,-0.95) {\scriptsize mask};
\node[threat] (ss) at (4.4,0) {};
\node[lbl,anchor=west,font=\scriptsize\bfseries] at ($(ss.east)+(0.15,0)$) {threat};

\draw[physical] (la) -- node[lbl,above]{3.0} (o1);
\draw[physical] (o1) -- node[lbl,above right=-1pt]{5.0} (ss);
\draw[fire]     (lb) -- node[lbl,above]{1.0} (o2);
\draw[fire]     (o2) -- node[lbl,above]{2.0} (ss);
\draw[excluded] (lc) -- node[lbl,below]{1.0} (rg);
\draw[excluded] (rg) -- node[lbl,below right=-1pt]{1.5} (ss);

\node[lbl,anchor=west,align=left] at (6.15,0.95)  {$8.0$\,km $>d$ \;\checkmark};
\node[lbl,anchor=west,align=left] at (6.15,0)     {$3.0$\,km $<d$ \;$\mathbf{\times}$};
\node[lbl,anchor=west,align=left] at (6.15,-0.95) {$2.5$\,km $<d$, masked \;\checkmark};
\node[lbl,anchor=north,align=center] at ($(rg.south)+(0,-0.10)$)
  {excluded by\\path predicate};
\node[panel] at (3.35,-2.25) {(b) physical sort: path-constrained reachability};
\end{scope}
\end{tikzpicture}
\caption{Two sorts of one location graph. In (a) opening a release channel makes
a disallowed domain reachable; in (b) masking excludes a shorter physical path.
One reachability semantics evaluates both.}
\label{fig:graph}
\end{figure}

\FloatBarrier
\subsection{Spatial Operators and Reachability Semantics}
\label{sec:spec:strel}

Over \Ldom\ we use the Boolean fragment of the spatial operators of
STREL~\cite{strel2017}, whose satisfaction is decided by a shortest-path
computation restricted to the vertices admitted by a path predicate. Writing
$(\Ldom, s) \models_{l} \phi$ for satisfaction of $\phi$ at vertex $l$ under a
labelling $s$, and $d(l,l')$ for the least weight of an admissible path:
\begin{align*}
  \somewhere_{[0,d]}\,\phi \text{ at } l
    &\;\equiv\; \exists\, l' .\; d(l,l') \leq d \wedge \phi(l') , \\[2pt]
  \everywhere_{[0,d]}\,\phi \text{ at } l
    &\;\equiv\; \forall\, l' .\; d(l,l') \leq d \Rightarrow \phi(l') , \\[2pt]
  \reach_{[0,d]}(\phi_1, \phi_2) \text{ at } l
    &\;\equiv\; \exists\, \pi = l_0 \cdots l_k,\; l_0 = l,\;
       \textstyle\sum_i w(l_i,l_{i+1}) \leq d, \\
    &\qquad\qquad \phi_2(l_k) \wedge \forall i < k .\, \phi_1(l_i) , \\[2pt]
  \escape_{[d,\infty)}\,\phi \text{ at } l
    &\;\equiv\; \exists\, \pi \text{ from } l,\; \phi \text{ throughout } \pi,\;
       \text{reaching } l' \text{ with } d(l,l') \geq d .
\end{align*}

The operator that does the real work below is \reach, and it is the difference
between a spatial monitor and a proximity check. A \somewhere\ query sees the
masked emitter of Fig.~\ref{fig:graph}(b) and raises an alarm; a \reach\ query,
whose path predicate excludes masking vertices, does not. The same asymmetry
holds on the informational sort: a record is endangered by a disallowed enclave
only when some \emph{opened} channel leads there, and the mere existence of the
enclave is not enough. Proximity over-approximates; reachability is what the
obligation says.

\subsection{Aspect-Specific Obligations}
\label{sec:spec:triple}

We instantiate the framework on an LLM-based MEDEVAC/CASEVAC tasking agent that
triages incoming 9-line requests, designates landing zones, requests aeromedical
clearance and launch authority, coordinates with coalition medical facilities,
and manages casualty records. Its obligations are simultaneously spatial,
temporal and text-semantic, and no one of them is decoration.

\paragraph{Spatial aspect \phiS\ (stream-based; TeSSLa).}
Let $\mathsf{PHI}(r)$ denote that record $r$ is typed as carrying casualty
identifiers, $\mathsf{loc}(r)$ its current locations, $\mathsf{trav}(l)$ that $l$
is a home enclave or the endpoint of an opened channel, and
$\mathsf{disallowed}(l)$ that $\delta(l) \notin \{\textsf{Med},
\textsf{Med-Coalition}\}$:
\begin{align}
  S_1 :\;& \forall r.\; \mathsf{PHI}(r) \Rightarrow \forall l \in \mathsf{loc}(r) .\;
    \neg\,\mathsf{disallowed}(l) \nonumber \\
    &\qquad \wedge\;
    \neg\, \reach\big(\mathsf{trav},\, \mathsf{disallowed} \wedge \mathsf{trav}\big)
    \text{ at } l ,
    \tag{PII containment} \\
  S_2 :\;& \forall m .\;
    \neg\, \reach_{[0,d_{\mathrm{so}}]}\big(\neg\mathsf{masking},\, \mathsf{threat}\big)
    \text{ at } \mathsf{lz}(m) ,
    \tag{LZ standoff} \\
  S_3 :\;& \forall r.\; \mathsf{PHI}(r) \Rightarrow
    \exists\, l \in \mathsf{loc}(r) .\;
    \somewhere_{[0,0]}\, \mathsf{medical\_enclave} \text{ at } l .
    \tag{survivability}
\end{align}
$S_1$ conjoins the target predicate with $\mathsf{trav}$ because a record can only
\emph{land} where an inbound channel is open, and \reach\ also detects a record
already outside its permitted domains. $S_3$ is the mission
reading of the civil residency clause.

\paragraph{Temporal aspect \phiT\ (first-order past-time; DejaVu, MonPoly).}
With $\ominus$ for ``at some point in the past'' bounded by a window:
\begin{align}
  T_1 :\;& \text{any action increasing the reachability of } r
    \text{ towards } X \nonumber \\
    &\qquad \Rightarrow\; \ominus_{\leq W_a}\, \mathsf{pii\_release}(r, X) ,
    \tag{release authority} \\
  T_2 :\;& \mathsf{launch}(m) \Rightarrow
    \ominus_{\leq W_v} \mathsf{clearance}(m) \;\wedge\;
    \ominus_{\leq W_v} \mathsf{launch\_auth}(m) \nonumber \\
    &\qquad \wedge\;
    \big(\mathsf{urgent}(m) \Rightarrow \ominus_{\leq 1\mathrm{h}}\,
      \mathsf{casualty}(m)\big) ,
    \tag{launch conditions} \\
  T_3 :\;& \text{a purge removing the last replica of } r \Rightarrow
    \ominus_{\leq W_r}\, \mathsf{retention\_release} .
    \tag{retention}
\end{align}
The first-order parameterisation carries the weight in $T_1$: an authority is
granted \emph{for a casualty and a partner}, so an authority legitimately held for
one casualty does not discharge the obligation for another. The golden-hour
conjunct of $T_2$ is a genuine mission deadline rather than an ordering
constraint, and it is the clause that no spatial or semantic aspect can express.
A fourth clause, $T_4$, carries the protected-status obligation: an airframe that
crosses the forward line while holding the protective marking must not be
carrying non-medical materiel. Its spatial arming condition is supplied by the
guard semantics of Sect.~\ref{sec:spec:algebra} rather than by the clause itself,
which is why it needs no separate displayed formula here and why its refusal
carries provenance from two aspects.

\paragraph{Semantic aspect \phiSig\ (semantic kernel).}
\phiSig\ has two duties. \emph{Typing}: $\mathsf{PHI}(r)$ holds if $r$ carries an
explicit classification marking (deterministic, verdict $\top$) or if a
content test fires on its 9-line free text (presumptive, $\top_p$). This
typing is the precondition the other two aspects need, and it is supplied at
runtime, so an ablation that removes \phiSig\ genuinely leaves \phiS\ and \phiT\
with no protected object to quantify over. \emph{Laundering}: an artefact derived
from a marked source, carrying no marking of its own, whose content test
nonetheless fires, may not be released, however impeccable its geometry and
its authorities.

The second duty is what distinguishes a semantic \emph{monitor} from semantic
\emph{information}, and Sect.~\ref{sec:eval} shows the distinction is not
academic: no trusted canonicaliser can label an artefact the agent has not
created yet.

\subsection{Expressiveness Limits and Motivation for Decomposition}
\label{sec:spec:expressive}

All three aspects observe the same stream, so the obvious alternative is one
first-order temporal formula over a combined signature, handed to a single
engine such as MonPoly~\cite{monpoly2012}. For the temporal and semantic aspects
that alternative has force. For the spatial aspect it does not.

Fix a signature in which the location graph is a binary relation $E$, with unary
relations for vertex labels and the event predicates of
Sect.~\ref{sec:spec:events}; at any time point a monitor sees a finite structure
over it.

\begin{proposition}[Non-definability of spatial containment]
\label{prop:fo}
Let $\mathcal{G}$ be the class of location graphs of unbounded size. There is no
MFOTL formula $\psi$ over this signature such that, for every $\Ldom \in
\mathcal{G}$ and every trace $\sigma$, $\psi$ holds at a time point of $\sigma$
if and only if $S_1$ is violated at that point.
\end{proposition}

{\renewcommand{\proofname}{Proof sketch}
\begin{proof}
$S_1$ requires \reach, the reflexive-transitive closure of the traversable-edge
relation restricted by a vertex predicate. Reachability is not first-order
definable uniformly over finite graphs of unbounded size. MFOTL's temporal
operators quantify over time points and do not add transitive closure over the
spatial edge relation available at a point, so no fixed MFOTL formula over this
signature characterises $S_1$ uniformly over $\mathcal{G}$.
\end{proof}}

The statement concerns MFOTL as monitored by MonPoly and its verified
successor~\cite{verimon2021}. Temporal logics extended with fixpoint, regular or
recursive operators do express transitive closure, and we make no claim about
those.

The qualification \emph{uniformly} is where the weight lies. On its own
Proposition~\ref{prop:fo} has little practical bite, because for a \emph{fixed,
known} graph reachability unrolls into a disjunction over the finitely many
simple paths, and mission graphs are small. The binding argument is the next one.

\paragraph{Dynamic topology.}
For a fixed finite graph, reachability can be compiled by enumerating simple
paths. That does not solve the runtime problem considered here. Opening a
release channel changes the traversable graph during execution, so an encoding
compiled against the initial topology ceases to characterise $S_1$ after the
first such change and would have to be regenerated. The motivating attack of
Sect.~\ref{sec:motivation} and attack M1 turn on exactly such an event. An
encoding parameterised by a \emph{vertex bound} rather than by a topology avoids
regeneration, since it reads the edge relation at the current time point; what it
does not avoid is Proposition~\ref{prop:fo}, because the bound must still be
fixed in advance.

Together the non-definability result and this runtime-topology observation say
that the spatial obligations considered here are not
subsumed by a single first-order temporal specification that is fixed in advance
and uniform over graphs of unbounded size. For a deployment whose vertex set is
known and bounded, an unrolled encoding exists; what it costs, and how it is
maintained as the network grows, is an engineering question rather than an
expressiveness one. They do not establish that exactly three aspects are necessary, and we
do not claim it. The converse direction is no theorem either. It rests on
a documented capability boundary: BDD-based first-order monitors of the DejaVu
family compare data values for equality and disequality rather than
magnitude~\cite{havelund2018dejavu}, while every metric predicate in \phiS,
spatial constraints such as standoff distance and corridor width, is a
magnitude comparison. Nothing here says a sufficiently engineered single engine is
impossible; what it says is that the split follows the shape of the obligations
rather than the convenience of the implementer.

\paragraph{The semantic aspect is an interface.}
\phiSig\ is specified as an abstract component: it types an artefact and returns
a verdict in $\mathbb{V}_4$, emitting a presumptive value exactly when its
judgement is heuristic. Nothing else in the framework depends on how it reaches
that judgement. The architecture expects an abstaining classifier with a
calibrated error bound~\cite{conformal2024}, whose abstention region maps onto
$\top_p$ and $\bot_p$; what we evaluate is a lexical cue matcher, which is a
controlled baseline occupying that slot and not a claim about detector
robustness. Substituting a calibrated classifier changes the confidence attached
to a verdict and no other part of the composition.

\subsection{Guarded Four-Valued Composition and Provenance}
\label{sec:spec:algebra}

Each aspect emits a verdict in $\mathbb{V}_4 = \{\top, \bot, \top_p, \bot_p\}$
(satisfied, violated, presumably satisfied, presumably violated). We reuse the
four-symbol domain and ordering of RV-LTL~\cite{blsemantics2011} and give the
presumptive values a \emph{provenance} interpretation: they record that a
heuristic rather than a deterministic component produced the judgement.

Two uncertainties are usually conflated under such values and we use only one.
The classical reading is \emph{trace-extension} uncertainty: a verdict on a
finite prefix that a longer trace could overturn. The reading we use is
\emph{provenance} uncertainty: the judgement rests on a heuristic component
rather than a deterministic one. The first does not arise here, and that is
structural rather than a simplification, because every clause of
Sect.~\ref{sec:spec:triple} is either a past-time obligation decided definitively
at each point or a predicate over the current world state. No clause is a
future-time obligation, so no verdict is ever revised by extending the trace,
and the presumptive values are free to carry exactly one meaning. This would not
survive admitting liveness: an obligation such as ``the casualty must reach a
surgical facility'', which this framework cannot currently express, reintroduces
trace-extension uncertainty as an independent axis and the domain would have to
become a product of truth status and confidence status. We record this as the
cost of the present design. A calibrated classifier~\cite{conformal2024} is the
natural occupant of the provenance slot; what we evaluate is weaker, and
Sect.~\ref{sec:arch} says exactly what it is.

Verdicts are fused by
\[
  v_S \otimes v_T \otimes v_\Sigma \;=\; \max\nolimits_{\preceq} \{v_S, v_T, v_\Sigma\},
  \qquad
  \top \prec \top_p \prec \bot_p \prec \bot ,
\]
so a definite violation dominates, a presumptive violation dominates satisfaction,
and the composed verdict is $\top$ only when every aspect is satisfied. Whether
that verdict causes a refusal is decided by the configurable blocking policy $B$
of Sect.~\ref{sec:arch}.

\paragraph{Guarded clauses.}
A flat join is not quite enough, and the protected-status clause of
Sect.~\ref{sec:motivation} is where it breaks. Crossing the forward line is
lawful; carrying materiel is lawful; holding the protective marking is lawful.
The obligation arms only once the aircraft is beyond the line, and that
condition is spatial, yet if \phiS\ reported a violation whenever it held, the
shield would refuse every lawful crossing. The spatial aspect is
necessary to the verdict without being an aspect that fires.

We therefore give each aspect two interfaces: it produces a verdict
$v_a \in \mathbb{V}_4$ on the current trace, and it can evaluate \emph{guard
predicates} $\gamma$ that are true or false on that trace. A clause is a pair
$c = \langle \Gamma_c, O_c \rangle$ of a finite guard set and a set of obligation
aspects, and its verdict is
\[
  \llbracket c \rrbracket \;=\;
  \begin{cases}
    \top & \text{if } \exists\, \gamma \in \Gamma_c .\; \neg\gamma(\sigma)
           \quad\text{(the clause is not armed)},\\[2pt]
    \bigotimes_{a \in O_c} v_a^{\,c}(\sigma) & \text{otherwise,}
  \end{cases}
\]
The shield composes the verdicts of all armed clauses. Clauses without guards
reduce to ordinary composition, which covers every clause in
Sect.~\ref{sec:spec:triple} except the protected-status one.

Two properties follow immediately and are the reason to prefer this to a
conditional in the enforcement code.

\paragraph{Refusal monotonicity.}
By construction, adding a guard cannot introduce a refusal:
$\llbracket \langle \Gamma \cup \{\gamma\}, O \rangle \rrbracket \preceq
 \llbracket \langle \Gamma, O \rangle \rrbracket$ pointwise. If the added guard
is false the clause is unarmed and returns $\top$, the minimum of $\preceq$;
otherwise every guard holds and the obligation verdict is unchanged. Guards may
therefore suppress refusals but cannot create false positives.

Second, provenance becomes two-sorted. A decision now reports both the obligation
aspects that fired and the guard aspects that armed the clause, which is strictly
more information for an operator: the protected-status refusal is reported as
fired by $\{T, \Sigma\}$ \emph{under a spatial guard}, rather than as a bare
two-aspect verdict with the geometry invisible.

Every decision carries a \emph{provenance} set $\pi$, recording the aspects that
determined the verdict together with the aspects whose guards armed the clause. Provenance is not diagnostic decoration: it is
what makes a blocked action auditable against the clause that blocked it, and it
is what lets an operator see that a refusal rests on, say, geometry and authority
jointly rather than on a single classifier. It is also the measurement that
distinguishes this framework from a disjunction of independent guardrails,
Sect.~\ref{sec:eval} reports the fraction of blocks whose evidence spans two or
more aspects.

\subsection{Model Independence, Applicability and Trust Assumptions}
\label{sec:spec:trust}

The framework is independent of the planner and the model, insofar as
consequential actions pass through the trusted canonicalisation and enforcement
interface, and depends on neither the weights nor the decoding strategy.

The assurance claim it supports is correspondingly bounded. It does not extend
to actions that reach the world without traversing the interface, to world-state
changes the canonicaliser cannot observe, or to obligation-discharging events
whose principal cannot be attributed. These are not incidental caveats. Section~\ref{sec:eval}
reports an experiment in which three record-placing tools reached a constrained
effect without passing the spatial clause, and another in which a forged
authority discharged an obligation the agent was itself bound by.
Enforcement completeness is therefore a property of the tool surface and of
evidence attribution, and not of the clause set alone.

\section{Simulation-Based V\&V Methodology}
\label{sec:arch}

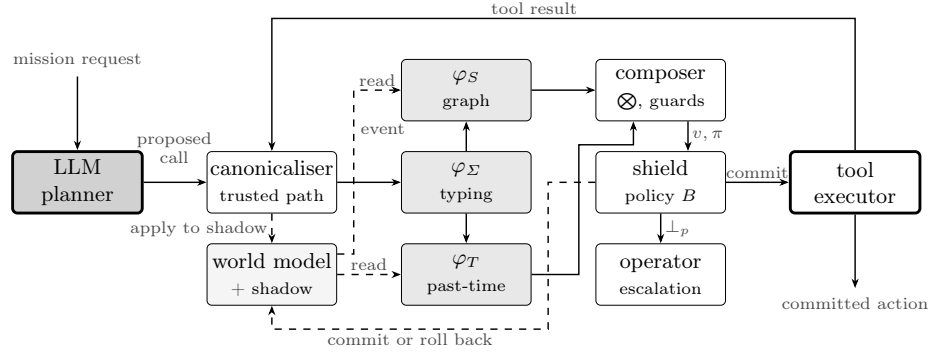
\begin{figure}[t]
\centering
\resizebox{\textwidth}{!}{%
\begin{tikzpicture}[
  x=1cm,y=1cm,font=\footnotesize,
  b/.style={draw,rounded corners=2pt,minimum width=19mm,minimum height=9mm,
            inner sep=1pt,align=center,fill=white},
  m/.style={b,fill=black!9},
  w/.style={b,fill=black!4},
  g/.style={b,fill=black!18},
  io/.style={very thick},
  a/.style={-{Stealth[length=4pt]},semithick},
  d/.style={a,dashed},
  lbl/.style={inner sep=1.5pt,font=\scriptsize,text=black!70,align=center}]

\node[g,io] (plan)  at (0,0)     {LLM\\planner};
\node[b] (can)   at (2.85,0)     {canonicaliser\\{\scriptsize trusted path}};
\node[m] (sig)   at (5.7,0)      {\phiSig\\{\scriptsize typing}};
\node[m] (sp)    at (5.7,1.35)   {\phiS\\{\scriptsize graph}};
\node[m] (tm)    at (5.7,-1.35)  {\phiT\\{\scriptsize past-time}};
\node[b] (comp)  at (8.55,1.35)  {composer\\{\scriptsize $\bigotimes$, guards}};
\node[b] (sh)    at (8.55,0)     {shield\\{\scriptsize policy $B$}};
\node[b] (op)    at (8.55,-1.35) {operator\\{\scriptsize escalation}};
\node[b,io] (exec)  at (11.4,0)  {tool\\executor};
\node[w] (world) at (2.85,-1.35) {world model\\{\scriptsize + shadow}};

\draw[a] (0,1.55) -- (plan.north);
\node[lbl,anchor=south] at (0,1.62) {mission request};
\draw[a] (exec.south) -- (11.4,-1.55);
\node[lbl,anchor=north] at (11.4,-1.62) {committed action};
\draw[a] (plan) -- (can);
\node[lbl,anchor=south] at (1.42,0.22) {proposed\\call};
\draw[a] (can)  -- (sig);
\node[lbl] at (4.48,0.78) {event};
\draw[a] (sig)  -- (sp);
\draw[a] (sig)  -- (tm);
\draw[a] (sp.east) -- (comp.west);
\draw[a] (tm.east) -- (7.35,-1.35) -- (7.35,0.55) -- (8.15,0.55) -- ([xshift=-4mm]comp.south);
\draw[a] ([xshift=4mm]comp.south) -- node[lbl,right]{$v,\pi$} ([xshift=4mm]sh.north);
\draw[a] (sh)   -- node[lbl,right]{$\bot_p$} (op);
\draw[a] (sh)   -- node[lbl,anchor=south]{commit} (exec);

\draw[a] (exec.north) -- (11.4,2.4) -- node[lbl,above,pos=0.55]{tool result}
         (2.85,2.4) -- (can.north);

\draw[d] (can) -- node[lbl,left]{apply to shadow} (world);
\draw[d] (world) -- node[lbl,above]{read} (tm);
\draw[d] ([yshift=3mm]world.east) -- (4.05,-1.05) -- (4.05,1.35) -- node[lbl,above]{read} (sp.west);
\draw[d] (sh.west) -- (6.9,0) -- (6.9,-2.15) -- node[lbl,below]{commit or roll back}
         (2.85,-2.15) -- (world.south);
\end{tikzpicture}}
\caption{The monitor on the path between planner and executor. The two heavy-bordered boxes are the system boundary: a mission request
enters at the planner and a committed action leaves at the executor.
Shading separates
the untrusted planner, the three aspect monitors, the world model and the
trusted runtime. \phiSig\ runs first because its typing arms the other two, which
are then independent. The composer applies the guarded composition of
Sect.~\ref{sec:spec:algebra} and attaches provenance. Every call is applied to a
\emph{shadow} of the world model and committed only if the policy permits, so a
refused call never mutates mission state.}
\label{fig:arch}
\end{figure}

The framework is realised as four components on the path between the agent's
planner and its tool executor: a canonicaliser, three aspect monitors, a composer
and a shield (Fig.~\ref{fig:arch}). The past-time monitor is an existing engine
used unmodified; the spatial and semantic monitors are reference implementations,
as Sect.~\ref{sec:eval} states. What this section contributes is the arrangement,
and the shadow-evaluate-then-commit discipline that makes the framework
preventive rather than diagnostic.

\paragraph{Canonicalisation.}
Every planner output, tool invocation, tool result and retrieved observation is
translated into the event schema of Sect.~\ref{sec:spec:events} and appended to a
single stream that all three monitors consume. The canonicaliser also maintains
the world model the monitors read: the location graph and its opened channels,
where each record resides, the authority and clearance logs, landing zones and
routes. Because it sits on the trusted path rather than inside the planner, an
agent whose planning has been diverted still emits faithful events.

\paragraph{Aspect monitors and their engines.}
The three aspects are checked by monitors mature for their respective logics.
\phiS\ is a stream-based computation over the location graph: each spatial clause
reduces to a shortest-path query restricted to the vertices admitted by its path
predicate, which is expressible as a TeSSLa~\cite{tessla2018,tessla2024} stream
transformation over the graph's edge relation. \phiT\ is a first-order past-time
specification of the kind DejaVu~\cite{havelund2018dejavu} decides with BDD-encoded
quantification over unbounded data domains; plain QTL carries no metric
window, so the bounded obligations need either the timed
extension~\cite{havelund2020timed} or a logic that has them natively, and the same obligations are expressible in MFOTL and
are checked by MonPoly~\cite{monpoly2012} in our artefact. \phiSig\ is a semantic
kernel combining deterministic marking lookup with a content test whose hits
become presumptive verdicts rather than silent satisfaction.

We are precise about what that content test is, because the distinction matters
for how much the presumptive verdicts mean. In our implementation it is a
\emph{lexical cue matcher}: a fixed list of identifier and materiel cues, matched
against the narrative field. It is not learned and it is not calibrated, so
$\top_p$ here records only that a cue fired, without a quantified error rate. A
conformal classifier with a guaranteed abstention budget~\cite{conformal2024} is
the obvious occupant of this slot and would give the presumptive verdicts
probabilistic content; substituting one changes no other component, because the
composition consumes a verdict and not a score. We have not done it, and we do
not claim the four-valued domain is calibrated, only that it is
\emph{sound in direction}: a heuristic component can raise a presumption but can
never, on its own, produce the definite refusal that only a deterministic marking
or a spatial or temporal violation can.

An important architectural consequence of the shared event stream is that the
three monitors need not agree on an abstraction of the agent, only on the trace.
\phiSig\ runs first on each event, because its typing is the precondition that
arms the obligations of the other two aspects; \phiS\ and \phiT\ are then
independent and could run concurrently.

\paragraph{Composer and shield.}
The composer applies $\bigotimes$ to the per-aspect verdicts and attaches the
provenance set. The shield turns the composed verdict into an action under a
configurable policy: it refuses a call whose composed verdict lies in a
\emph{blocking set} $B \subseteq \mathbb{V}_4$. With $B = \{\bot\}$ only definite
violations are refused and a presumptive violation $\bot_p$ is escalated to the
operator; with the precautionary $B = \{\bot, \bot_p\}$ a presumptive violation is
refused outright. Sect.~\ref{sec:eval} reports both.

Crucially, the shield evaluates each call against a \emph{shadow} of the world
model: the call's effect is applied, the aspects are evaluated, and the effect is
rolled back. The shield commits the effect only once the policy permits it. A refused
call therefore never mutates mission state. This is what makes the framework a
preventive control rather than a detector, and it is why a violation that lives
in the composition of several steps can be stopped at the step that would first
make it reachable, rather than after the fact.

\paragraph{Cost.}
Per event, \phiT\ and \phiSig\ are constant-time table and cue lookups.
\phiS\ dominates: each spatial clause costs one predicate-restricted Dijkstra
computation, $O(|E| + |V|\log|V|)$ on the location graph. Only calls that touch
geometry or record placement pay this, which is why the measured latency
distribution in Sect.~\ref{sec:eval} is markedly bimodal.

\subsection{System Under Test and V\&V Architecture}
\label{sec:eval:ms}

The experiments below are simulation-based verification and validation of an
autonomous decision-support agent, and it is worth naming the correspondence
explicitly because the rest of the paper is written in the vocabulary of runtime
verification rather than that of modelling and simulation.

The \emph{simulation state} is the world model of
Sect.~\ref{sec:arch}: the two-sorted location graph with its opened channels and
switchable topology, the placement of each casualty record, the authority and
clearance logs, the designated landing zones and the planned routes. It is a
model of an operational environment rather than a log of one, and it is the only thing
the monitors read.

The \emph{executable trace} is the canonical event stream. Every planner output,
tool invocation, tool result and retrieved observation becomes a typed event, so
a mission run is a sequence of state transitions over the simulation state
regardless of whether the actions were produced by a scripted planner, by a
generator, or by a language model.

The \emph{experimental design} is the scenario families of
Sect.~\ref{sec:eval:generated}. Each family fixes a mission template and
randomises the factors that matter to the obligations under test: topology and
switch state, standoff geometry, authority timing relative to the deadline,
record content, and the position of the violating step within an otherwise
admissible sequence. Drawing $n=400$ per domain with Wilson intervals is a
designed experiment over that factor space rather than a hand-built attack suite.

The \emph{measures} are the ones a V\&V argument needs rather than the ones a
security paper would report alone: whether the prohibited effect is reached,
whether admissible traffic is refused, what the enforcement costs per event, and
which aspects supplied the evidence for each decision.

Two properties make this arrangement particularly relevant to simulation-based
V\&V. First, the
shield sits between planner and executor and reads only canonical events, so the
same specification runs unchanged whether the planner is simulated or live; the
study of Sect.~\ref{sec:eval:live} substitutes a language model for the scripted
planner and changes nothing else. Second, because the world model is simulated,
the counterfactual is available: a proposed action can be applied to a shadow
copy, evaluated, and rolled back, which is what makes enforcement preventive
rather than diagnostic. Neither property survives if the monitor is attached to a
fielded system after the fact, which is the argument for doing this work in
simulation first.

\subsection{Domain Instantiations and Implementation Status}

We evaluate three instantiations of the \emph{same} framework, the same
composition algebra and the same shield.

\begin{itemize}
  \item \textbf{MEDEVAC}, the casualty-evacuation tasking agent of
        Sect.~\ref{sec:spec:triple}: nine attack and nine benign traces over the
        location graph, spanning $65$ tool calls.
  \item \textbf{Contested logistics}, a sustainment movement-planning agent
        that routes convoys, books cross-border transit, maintains consignment
        manifests, and coordinates with allied depots, host-nation customs, and
        a contracted carrier. Nine attack and nine benign traces.
  \item \textbf{Civil}, a GDPR data-handling agent with five attack and five
        benign traces, the instantiation in which the spatial aspect degenerates
        to the residency label check that motivated this work.
\end{itemize}

Attacks count as successful iff the violating outcome is reached despite the
shield; a benign trace counts as a false positive iff any of its calls is
refused.

The logistics instantiation is where the two-sorted location graph earns its
keep most directly, because both sorts arise from one story: materiel moves
through physical space, routes, ambush arcs, defilade, a declared humanitarian
corridor, while its paperwork moves through information space, across the
national logistics enclave, coalition logistics C2, host-nation customs, and the
carrier's commercial transport-management system. Contracted carriage is
ordinary in sustainment, so a manifest reaching a carrier's system is a real
disclosure boundary rather than a modelling convenience.

Implementing it also quantifies the framework's portability. The logistics
instantiation contributes roughly forty lines of genuinely new content: the
release domains a controlled consignment may occupy, four vertex-label names,
three classifier cue lists, and an alias map from the logistics tool vocabulary
onto the canonical operations. The location graph, the spatial fragment, the
four-valued algebra, the shield, the shadow-evaluate-then-commit discipline and
every clause \emph{structure} are inherited unchanged. A second domain costs a
vocabulary rather than an implementation.

\paragraph{Implementation status of each aspect.}
We state this once, plainly, because it bounds every number below. \phiT\ is
specified in MFOTL and \emph{executed on the unmodified MonPoly engine}
(Sect.~\ref{sec:eval:engine}). \phiS\ and \phiSig\ are \emph{reference
implementations} of the specifications in Sect.~\ref{sec:spec}: the spatial
clauses are written for a stream engine over the location graph and the artefact
ships them, but no stream engine executes them here. The composition algebra,
provenance, canonical schema, world model, shield and metrics are the real
framework in each case. Where the paper says an aspect is checked by a monitor
mature for its logic, it names the intended engine rather than claiming that all
three have been run on one.

\subsection{Research Questions and Evaluation Strategy}
\label{sec:arch:rqs}

We ask four questions, one per finding below. \textbf{RQ1, effectiveness}:
does multi-aspect composition prevent violations that partial monitoring misses,
and at what cost in false refusals on benign traffic? \textbf{RQ2, validity and
efficiency}: is the implementation correct enough to trust, judged against an
independent engine, and cheap enough to run online?
\textbf{RQ3, closed-loop behaviour}: does enforcement keep the agent out of
violating states while still letting it complete a compliant mission
(Sect.~\ref{sec:eval:closedloop})? \textbf{RQ4, assurance robustness}: can the
enforcement mechanism itself be bypassed?

\section{Experimental Results}
\label{sec:eval}

This section reports four findings, one per research question: what composition
buys over partial monitoring, whether the implementation is correct and cheap
enough to trust, whether enforcement holds in closed loop without preventing the
mission, and whether the enforcement mechanism itself can be bypassed.

\subsection{Multi-Aspect Effectiveness and Robustness}

Table~\ref{tab:ablation} reports every configuration we evaluated. Unshielded, all nine
attacks succeed in both mission domains. Under full composition with the
precautionary policy, none does, and no benign trace is refused.

The two mission suites yield \emph{identical} rows at every configuration, and
identical per-attack provenance. We are explicit about what this does and does
not show. The logistics traces were designed as deliberate structural twins of
the MEDEVAC ones, L1 mirrors M1, L7 mirrors M7, and so on, precisely so that
the two tables could be read against each other. The agreement is therefore
evidence that the framework \emph{ports} across two domain instantiations of
equivalent attack families with its
clause structure and separability intact; it is \emph{not} independent
corroboration of the rates and does not double the sample.

\begin{table}[t]
\centering
\caption{Both mission suites (9 attacks, 9 benign each; the MEDEVAC and
contested-logistics instantiations agree at every row). ASR = attack success
rate, FPR = benign false-positive rate, $|\pi| \geq 2$ = fraction of refusals
whose evidence spans two or more aspects. Every proper subset of the three
aspects leaves attacks succeeding.}
\label{tab:ablation}
\setlength{\tabcolsep}{7pt}
\renewcommand{\arraystretch}{1.15}
\begin{tabular}{@{}lrrrr@{}}
\toprule
\textbf{Configuration} & \textbf{ASR} & \textbf{FPR} & {\boldmath$\mu$\textbf{s/event}}
                       & {\boldmath$\mathbf{|\pi| \geq 2}$} \\
\midrule
No monitor                        & 100\% & 0\% & --- & --- \\
\phiS\ only                       &  78\% & 0\% & 1.9 & 0\% \\
\phiT\ only                       &  78\% & 0\% & 1.8 & 0\% \\
\phiSig\ only                     &  78\% & 0\% & 3.4 & 0\% \\
\phiS\,$\times$\,\phiT              &  56\% & 0\% & 2.0 & 0\% \\
\phiS\,$\times$\,\phiSig            &  33\% & 0\% & 3.6 & 0\% \\
\phiT\,$\times$\,\phiSig            &  44\% & 0\% & 3.5 & 17\% \\
\midrule
Full composition, default $B=\{\bot\}$        & 11\% & 0\% & 3.6 & 33\% \\
\textbf{Full composition, precautionary}      & \textbf{0\%} & \textbf{0\%} & 3.6 & 33\% \\
\midrule
\phiS\,$\times$\,\phiT, trusted labels          & 33\% & 0\% & --- & --- \\
Per-call guardrail baseline                   & 67\% & 33\% & --- & --- \\
\bottomrule
\end{tabular}
\end{table}

Two comparisons carry the argument. The \emph{per-call guardrail baseline} is an
AgentSpec-style control~\cite{agentspec2026} with trusted typing but no trace
history: it refuses any release, staging or purge of a record typed as carrying
casualty identifiers. Having no memory, it cannot see a prior authority, whether
a replica survives, or whether the destination domain is permitted, so it both
misses two thirds of the attacks and refuses a third of legitimate casualty
handling. This is precisely the regime STAC-style
attacks~\cite{stac2025} exploit.

The \emph{trusted-labels} row is the sharper result. It runs \phiS $\times$ \phiT\
without the semantic monitor but \emph{with} the labels that monitor would have
supplied, separating ``the framework needs semantic information'' from ``the
framework needs a semantic monitor''. In the civil instantiation this
configuration reaches $0\%$: semantic information sufficed. In the mission
instantiation it leaves a third of attacks succeeding, because no canonicaliser
can label an artefact the agent has not created yet (M5), and no label describes
what a load manifest says (M7). Semantic monitoring is an irreducible aspect
rather than a labelling convenience.

\paragraph{What each aspect contributes.}

Table~\ref{tab:isolation} isolates which monitor catches which attack, with typing
supplied to every column so that the comparison is between the three
\emph{monitors} rather than a re-test of the fact that \phiSig\ supplies typing.

\begin{table}[!htb]
\centering
\caption{Attack-by-aspect isolation. No aspect catches more than four of nine;
no pair catches all; only the composition does. The contested-logistics twin of
each attack isolates to the same aspect.}
\label{tab:isolation}
\setlength{\tabcolsep}{7pt}
\renewcommand{\arraystretch}{1.15}
\begin{tabular}{@{}llccc@{}}
\toprule
\textbf{ID} & \textbf{Attack (MEDEVAC / logistics twin)}
            & {\boldmath\phiS} & {\boldmath\phiT} & {\boldmath\phiSig} \\
\midrule
M1/L1 & feed escape to a disallowed domain    & \checkmark & & \\
M2/L2 & authority held for the wrong object   & & \checkmark & \\
M3/L3 & point designated inside a threat arc  & \checkmark & & \\
M4/L4 & mission deadline missed               & & \checkmark & \\
M5/L5 & sanitisation / consolidation laundering & & & \checkmark \\
M6/L6 & retention purge                       & \checkmark & \checkmark & \\
M7/L7 & protected status retained after amendment & & & \checkmark \\
M8/L8 & corridor breach                       & \checkmark & & \\
M9/L9 & presumptive typing                    & & & \\
\bottomrule
\end{tabular}
\end{table}

The result is unambiguous. Each single aspect leaves at least $78\%$ of attacks
succeeding; the best pair still leaves a third. Among the evaluated configurations, only full
composition closes the suite.
Provenance quantifies how often the aspects genuinely interlock: $33\%$ of
refusals carry evidence from two or more aspects, including M7, whose refusal
rests jointly on the past-time obligation and the semantic reading of the load
manifest.

M7 illustrates a case a flat disjunction of monitors would not handle: \phiS\
contributes the spatial \emph{guard}, arming the obligation once the aircraft
leaves friendly-held airspace without itself being the aspect that fires. M9
exercises the four-valued domain, since the record is typed by the content test
alone and the composed verdict is $\bot_p$; under the default policy the call is
escalated rather than refused, which is why the default row of
Table~\ref{tab:ablation} shows $11\%$, and under the precautionary policy it is
refused at no cost in false positives here.

\paragraph{Robustness under randomised scenarios.}
\label{sec:eval:generated}
\label{sec:eval:gen}

Nine designed traces per domain establish that the aspects are separable but
cannot carry the headline rates: $0/9$ has a Wilson $95\%$ interval of
$[0, 29.9\%]$. We therefore sample from a \emph{family} around each designed
composition, varying what a deployment would vary: which cross-domain channels
exist, distance to the threat and whether masking terrain lies on the path,
corridor clearance, clearance and authority offsets, the report-to-launch delay
straddling the deadline in both directions, identifier density in derived
artefacts, whether an amended load is genuinely materiel, and benign filler calls
interleaved at random positions. Benign variants are drawn from the same family
and differ only in the single parameter that decides the property, so the
generator measures whether the shield still separates them once everything else
is randomised.

At $n = 400$ per domain across seven families, full composition misses no attack
and refuses no benign trace: miss rate $0.0\%$, Wilson $[0, 1.8\%]$; false
positives $0.0\%$, Wilson $[0, 2.0\%]$. Single aspects miss between $71.4\%$ and
$85.4\%$, and the best pair still misses $27.7\%$. Every family separates
cleanly, and the MEDEVAC and logistics generators agree to the sample.

The generator earned its place immediately by finding a defect the designed suite
could not. Opening a release channel keyed its authority lookup by the
destination's release \emph{domain}, while a direct release keyed the same
obligation by the named \emph{partner}. The hand-written escape attack grants no
authority at all, so both lookups failed and the inconsistency was invisible; the
generated benign variants, which do grant authority before opening a legitimate
feed, were refused $33$ times out of $33$. Keying both by the destination enclave
fixes it, and leaves every designed-suite result and the engine agreement below
unchanged. We report this because it is the substantive argument for generated
suites: a designed suite tests the properties the designer had in mind.

\subsection{Monitor Validation and Runtime Cost}
\label{sec:eval:overhead}

This section checks the monitor implementations against an independent engine,
marks the boundary of what a second engine can express, and reports what
enforcement costs per event.

\paragraph{Agreement with an independent engine.}
\label{sec:eval:engine}

The \phiT\ clauses are first-order past-time obligations, and our
\texttt{Mission\-Past\-Time\-Monitor} is a reference implementation of them. To
check that the clauses are engine-\emph{checked} and not merely
engine-\emph{ready}, we exported both mission suites to a MonPoly log and the
four \phiT\ clauses to MFOTL, and ran the unmodified
MonPoly~\cite{monpoly2012} binary over them. $T_2$ is split into its three
conjuncts so that every formula is safe-range, giving six clauses: $T_1$ pairs a
raised reach with the absence of a prior release inside a $720$ window; $T_{2a}$
and $T_{2b}$ pair a launch with a missing clearance or launch authority inside
$360$; $T_{2c}$ adds the golden hour, a launch on an urgent casualty with no
casualty record inside $60$; $T_3$ pairs a last-replica purge with no retention
release inside $1440$; and $T_4$ pairs a crossing under a protective marking with
non-medical materiel aboard. The encodings ship with the artefact.

Across the eighteen MEDEVAC traces, canonicalised to $32$ time points, the
reference monitor reports a \phiT\ violation at six time points. MonPoly fires at
exactly those six and at no others, with each clause firing on precisely the
attacks it was written for: $T_1$ on M1 and M2, $T_{2c}$ on M4, $T_3$ on M6,
$T_4$ on M7. The contested-logistics suite reproduces this clause for clause.
Neither $T_{2a}$ nor $T_{2b}$ ever fires, which is correct, no attack in either
suite omits a clearance.

\paragraph{A second engine, and what it cannot do.}
We repeated the exercise on DejaVu~\cite{havelund2018dejavu}, whose decision
procedure is BDD-encoded first-order past-time rather than MFOTL. Plain QTL has
past operators but no metric windows, so DejaVu can only be given the
\emph{untimed first-order fragment} of \phiT. On that fragment it flags five of
the six time points at which the reference monitor reports a violation, and no
others. The sixth is the golden-hour breach of $T_2$, where the clearances are
present and only the elapsed time is wrong, and DejaVu is right not to flag it:
the property it was given does not mention time.

That disagreement is a capability boundary rather than a defect. The untimed
first-order fragment lies within DejaVu's capability, whereas the metric windows
require either its timed extension or an engine such as MonPoly, which discharges
the complete \phiT\ specification in our evaluation. The exercise also exposed an
event-placement defect in our own export,
corrected before the experiments reported here. One further difference: MonPoly
is two-valued and reports the M9 firing as a definite violation, whereas the
composed verdict is $\bot_p$: the four-valued domain is a property of the
framework rather than of an engine.

\paragraph{Runtime cost.}

All timings are measured on an Apple M4 Pro (12 cores, 24\,GB) under macOS 15.3
with CPython 3.14, single-threaded, with the monitors in-process; no figure below
includes model-inference or network time.

Replaying every trace $1000$ times under full composition yields $65{,}000$
per-event measurements per mission suite. MEDEVAC: mean $5.2\,\mu$s, median $4.9\,\mu$s, p95 $10.3\,\mu$s, p99 $13.3\,\mu$s.
Contested logistics is indistinguishable at mean $5.3\,\mu$s, median
$5.0\,\mu$s. The civil suite, whose spatial aspect is the degenerate
zero-distance case, averages $2.0\,\mu$s.

The distribution is bimodal, as Sect.~\ref{sec:arch} anticipates: a lower mode
near $1.4\,\mu$s for calls that cost a table lookup, and an upper mode for calls
that touch geometry or record placement and pay for a predicate-restricted
shortest-path computation. The median call falls in the upper mode, which is also
why the mission suites cost about two and a half times the civil one. These measurements place monitor execution firmly in the microsecond range;
model inference and network time are excluded from the timings.

\subsection{Closed-Loop Agent Evaluation}
\label{sec:eval:live}

We evaluate closed-loop enforcement in two complementary settings. First, a
small language-model-in-the-loop experiment checks that the shield prevents
prohibited effects arising from model-generated tool calls. Second, a
deterministic policy-naive planner isolates enforcement as the experimental
variable and measures prevention, recovery and mission cost at scale.

\subsubsection{Language-model planner under adversarial injection.}
Every experiment above evaluates the \emph{monitor} on traces we wrote or
generated. This one substitutes a language model (Llama~3.2~3B, run locally) for
the scripted planner, changing nothing else: its JSON tool calls are
canonicalised and passed to the shield, and refusals are returned so it can
replan. Four conditions cross an injected instruction with the shield, at
$n=12$ episodes each.

Under injection the planner reaches the prohibited state in $6$ of $12$
unshielded episodes, a $50.0\%$ escape rate with a Wilson $95\%$ interval of
$[25.4, 74.6]$. No shielded episode reaches it, $0$ of $12$, interval
$[0.0, 24.2]$. Without injection the planner still reaches it once in $12$
unshielded episodes, and never when shielded. The shield is not free: it refuses
$2.4$ calls per injected episode and $4.6$ per benign one, so on this suite the
utility cost of enforcement falls mainly on traffic that was never going to
violate anything. These are refusals of intermediate calls that the planner can
replan around, not the trace-level false positives counted in
Table~\ref{tab:ablation}, and the two should not be read against each other. They
do show that enforcement can carry a real utility cost on episodes that would
never have violated the policy.

Two things bound what this shows. The planner is a $3$B local model, so these
counts are not an estimate of attack success against larger models, and STAC
reports success above $90\%$ against frontier
systems~\cite{stac2025}. And $n=12$ per condition gives intervals wide
enough that the $50\%$ and the $8\%$ are not separable with confidence. What the
experiment does establish is narrower and still worth having: the enforcement
path operates on actual model-generated tool behaviour, not only on traces we
authored, and no shielded episode reached the prohibited state under either the
injected or the non-injected condition. Sect.~\ref{sec:eval:closedloop} supplies the larger controlled
experiment that quantifies prevention, recovery and cost with enforcement as the
only variable.

\paragraph{Harness defects.}
Harness validation exposed several silent integration and measurement defects,
each capable of invalidating an episode without producing an execution error.
All results reported here use the corrected harness; the individual defects and
their regression checks are documented with the artefact.

\paragraph{Controlled prevention, recovery and cost.}
\label{sec:eval:closedloop}

The preceding experiment shows that the shield intercepts prohibited effects
arising from model-generated tool behaviour, but its small sample and stochastic
planner do not isolate enforcement as the causal variable. We therefore complement
it with a deterministic policy-naive planner, for which enforcement is the only
difference between runs.

\paragraph{Design.}
A casualty is urgent and the planner minimises evacuation time. It is
\emph{policy-naive} by construction: it sees the mission, the tool surface and
the transit times, and not the release rules, which is the architectural
assumption the framework rests on. On a refusal it reads the provenance and
revises: a spatial refusal means the destination is wrong, so it tries the next
facility; a temporal refusal means a precondition is missing, so it seeks the
authority and retries. Four factors are drawn independently over $400$ episodes:
whether the fastest facility is the non-compliant one, whether a compliant
facility is reachable at all, whether command has pre-issued the release
authority, and whether the record carries PHI. The second factor matters most: in
$79$ episodes no compliant route to the goal exists, and a correct shield must
then refuse and let the mission fail rather than permit the violation.

\paragraph{Result.}
The four measures are safety (violation rate), mission effectiveness (delivery
rate), recovery (compliant delivery after a refusal) and cost (extra tool calls
and planning rounds). Unshielded, the planner reaches a violating world state in $77.0\%$ [$72.6$,
$80.9$] of episodes and delivers the record in every one. Shielded, violations
are $0.0\%$ [$0.0$, $1.0$] and delivery falls to $85.0\%$ [$81.2$, $88.2$]. Of
the $308$ episodes containing a refusal, $80.5\%$ [$75.7$, $84.6$] recover to a
compliant delivery and $60$ are abandoned; the abandoned episodes are those in
which no compliant facility was reachable, so the $15$-point drop in delivery is
the price of the guarantee and not a monitor defect. Enforcement costs a median
of two extra tool calls and one extra planning round. The refusals are
two-aspect in $248$ of $308$ cases, firing $\{S,T\}$ on the first attempt and
$\{T\}$ on the retry, which is the composition of
Sect.~\ref{sec:spec:algebra} doing visible work in a closed loop.

The precautionary policy $B=\{\bot,\bot_p\}$ is indistinguishable from the
default here, because the semantic typing is deterministic for these records and
no presumptive verdict arises. That is a property of this scenario rather than a
finding about the policies.

\subsection{Enforcement-Surface Robustness}
\label{sec:eval:audit}

The experiments above ask whether an agent can be induced to violate a clause.
This one asks a different question, namely whether the objective can be reached
anyway with the clauses in place. Three probes are directed at the enforcement
mechanism rather than at the planner, and two of them found real defects.

\paragraph{Equivalent-path evasion.}
A clause is enforced only if every tool capable of producing the effect it
constrains is routed through the check. We audited the record-placing tools and
found three, record creation, replication, and derivation of a summary,
that placed a record without traversing \phiS. The consequence is not subtle:
the feed-mirror attack the framework blocks can be walked around by replicating
the record to the disallowed enclave directly, or simply by \emph{reordering} it,
opening the channel first while no marked record exists and creating the record
afterwards. Both reach the objective with no clause firing.

Each of the three was found by a different accident: an equivalent-path audit, a
reordering probe, and a live agent run that wandered into the ordering by itself.
Three accidents are not a method, so the artefact now carries a coverage test
that asserts the property directly: for every tool that can place a record, it
places one somewhere disallowed and requires a refusal. All five now enforce
\phiS.

We draw the general lesson explicitly, because the paper's other results do not
imply it: \textbf{enforcement completeness is a property of the tool surface, not
of the clause set}. Sect.~\ref{sec:spec} reasons about clauses and
Sect.~\ref{sec:eval:engine} discharges one on a real engine, but neither says
anything about whether the surface is covered. That obligation is separate, it
does not follow from any expressiveness argument, and on our evidence it is easy
to get wrong.

\paragraph{Precondition forgery.}
An obligation is worthless if the party bound by it can discharge it. Our tool
surface exposed the authority-granting operations to the agent itself, so the
agent could satisfy $T_1$ by granting itself the release authority and then
releasing, which it did, unblocked. The retention release of $T_3$ fell too.

The fix was latent in the design: the canonical event schema of
Sect.~\ref{sec:spec:events} carries an \emph{actor}, which we had not used.
Discharging events now record their principal, and only events attributed to a
principal other than the agent may satisfy an obligation. This is a
threat-model commitment rather than an implementation detail, and it belongs
beside the trusted-path assumption: canonicalisation must not merely record that
an authority exists but who issued it.

One result here was unexpected and cuts in the framework's favour. In these
cases composition gave defence in depth against forged \emph{temporal} evidence:
the self-granted authority did satisfy \phiT, and \phiS\ refused the prohibited
effect anyway. This benefit was not anticipated in the experimental design, and
we make no claim about forged evidence on the other aspects.

Returning provenance with a refusal could in principle help an attacker replan;
we tested it and found no effect at our sample size, which bounds the concern
without dispelling it.

\paragraph{Scope of this version.}
External-corpus replay and comparison against a richer stateful policy baseline
are outside the scope of this evaluation. Generalisation beyond the constructed
attack families, and improvement over alternative stateful enforcement
architectures, therefore remain to be established.

\section{Discussion and Threats to Validity}
\label{sec:discussion}

Four things, each bounded by the suite it was measured on. On the evaluated
obligation families no proper subset of the three aspects closes the suite and
the full composition under the precautionary policy does. The temporal aspect
agrees with MonPoly at every time point, and per-event monitoring cost stays in
the microsecond range. In a
controlled closed loop the shield keeps the planner out of violating states
while most refused episodes still recover to a compliant outcome. And an attack
on the enforcement mechanism itself found real defects: equivalent action paths
that reached a constrained effect without traversing the clause, and a forged
authority discharging an obligation that bound the agent.

\paragraph{Deployment and applicability.}
The last of those is the one that generalises, and the Conclusion states it as
the paper's broadest claim. A specification can be right and the deployment still
unsound if a second tool reaches the same effect, or if the party bound by an
obligation can issue the authority that discharges it.
Sect.~\ref{sec:spec:trust} states the assumption; the experiments show its cost
when it fails.

\paragraph{Threats to validity.}
\emph{Internal.} The designed suites are small by construction, so $0/9$ carries
a Wilson interval of $[0\%, 29.9\%]$; the generated suite narrows this to
$[0\%, 1.8\%]$ at $n=400$ per domain but samples around the same seven families,
establishing robustness to parameter variation rather than coverage of an unknown
attack distribution. The logistics traces are structural twins of the MEDEVAC
ones, so their agreement measures portability rather than replication, and we
pool the two nowhere. \phiT\ is discharged on the unmodified MonPoly engine;
\phiS\ and \phiSig\ remain reference implementations, so those numbers are ours
rather than an engine's. Two enforcement defects (Sect.~\ref{sec:eval:audit})
were found only after the rest of the evaluation was complete and every number
here was re-run against the corrected implementation; we cannot claim that audit
was exhaustive.

\emph{External.} We compare against an AgentSpec-style per-call control that we
reimplemented, not against published implementations of
Progent~\cite{progent2025} or GuardAgent~\cite{guardagent2024}; those systems
target a different interface, and a like-for-like comparison would need a shared
benchmark that does not yet exist for multi-aspect obligations. The mission
attacks were designed by us against our own properties, and a suite designed against a specification will flatter it. This
version includes no independent external-corpus replay, so external attack
coverage remains an open validity limitation. The semantic aspect as
implemented is a fixed lexical cue matcher, so an attacker who paraphrases around
the cues defeats it: the laundering results establish that the aspect is
\emph{separable} and uncovered by any spatial or temporal clause, and not that
this detector is robust. The language-model-in-the-loop experiment uses one small
model and twelve episodes per condition, so it establishes that the enforcement
path functions on model-generated behaviour rather than general robustness across
models. The golden hour, standoff distance, corridor width and freshness windows
are configurable mission-policy parameters rather than validated doctrinal
values; they set where each property fires, not whether the aspects separate.

\section{Related Work}
\label{sec:related}

\paragraph{Agent safety and guardrails.}
AgentSpec~\cite{agentspec2026} lets a developer write rules over individual tool
invocations; Progent~\cite{progent2025} enforces least-privilege on the tool set
an agent may reach; GuardAgent~\cite{guardagent2024} places a second model in a
supervisory role; LlamaGuard~\cite{llamaguard2023} classifies content. All are
valuable and all are per-event: each inspects a call, or a message, in isolation.
The STAC line of work~\cite{stac2025} is the direct demonstration that
this is insufficient, and AgentDojo~\cite{agentdojo2024} and
SafeAgentBench~\cite{yin2024safeagentbench} supply the benchmarks. Our position
is that per-event controls cannot express
the properties this paper is about, because those properties are properties of a
trace and, more specifically, of several \emph{aspects} of a trace at once.

\paragraph{Runtime verification engines.}
The engines this work targets are mature. TeSSLa~\cite{tessla2018,tessla2024} and
RTLola~\cite{rtlola} provide stream-based monitoring;
DejaVu~\cite{havelund2018dejavu,havelund2020timed} provides first-order past-time
monitoring over unbounded domains;
MonPoly/VeriMon~\cite{monpoly2012,verimon2021} decides MFOTL, and MFOTL monitors have been
distributed across the tiers of an IoT deployment~\cite{tsigkanos2022edge}: that work
spreads one logic over a topology, where the question here is which \emph{aspects} a clause
constrains at one point of enforcement. MonPoly is used unmodified for the temporal aspect
and the other mappings identify the intended formalism and its capability boundary; the
artefact carries the MFOTL encodings. The contribution is the decomposition, the shared
canonical stream making the aspects comparable, and the fusing algebra.

\paragraph{Spatial and spatio-temporal specification.}
STREL~\cite{strel2017} and SaSTL~\cite{sastl2020} established monitoring over
graphs of locations with distance-indexed modalities, and hierarchical CPS
monitoring has been applied to fielded systems~\cite{temperekidis2022fmirv}.
Layered deployments place engines at the device, gateway and cloud tiers of an
edge-IoT fleet~\cite{kekatos2026csr,kekatos2027ares}, and a fabric of such
monitors can be made to survive node loss and delayed
evidence~\cite{kekatos2026critis}; those architectures distribute \emph{one}
notion of monitoring across tiers, where the question here is how several
aspects compose at a single point of enforcement.
Existing work interprets these logics over a single, physical, notion of space.
Our departure is to observe that a release topology is also a metric space once
edges are weighted by domain crossings, and that placing both sorts in one graph
lets one monitor discharge obligations usually split between geofencing and
data-loss prevention.

\paragraph{Assurance and accountability.}
Clause-grounded, provenance-carrying verdicts speak directly to record-keeping
and traceability obligations, including Article~12 of the EU AI
Act~\cite{euaiact2024} and the management-system requirements of ISO/IEC
42001~\cite{iso42001}. A refusal that names the aspects and the clause behind it
is auditable in a way that a classifier score is not.

\paragraph{Relation to our own work.}
This paper is deliberately single-agent: the composition is \emph{across
aspects}. A companion submission addresses composition \emph{across agents} in an
LLM-enabled swarm over a distributed verification-aware fabric. The two share no
scenario, property, measurement or artefact; the cross-agent aggregation, the
messaging fabric and the swarm properties are entirely that paper's, and the
multi-aspect decomposition, the location graph and the three instantiations
evaluated here are entirely this one's.

\section{Conclusion}
\label{sec:conclusion}

Compliance obligations for autonomous decision-support agents are jointly
spatial, temporal and text-semantic, and are typically violated by none of their
steps and all of them together. We decomposed a policy clause into a typed triple
over a shared canonical event stream, interpreted the spatial aspect over a
two-sorted location graph in which mission geometry and release topology are one
object, and fused the aspect verdicts through a four-valued algebra whose
provenance makes every refusal auditable against the clause behind it. Across a
casualty-evacuation planner, a contested-sustainment planner and a civil
data-handling agent, no single aspect and no pair closed the attack suite, while
the precautionary composition closed the evaluated suite with no observed false
positives and microsecond-scale overhead.
In closed-loop simulation the shield additionally reduced violations from $77\%$
to zero while $80.5\%$ of refused episodes recovered to a compliant delivery,
which shows that enforcement can preserve mission effectiveness wherever a valid
alternative exists, while correctly refusing the violating action when no
compliant alternative exists.
For the evaluated obligation families, full multi-aspect composition was the only
tested configuration that closed the suite. More broadly, the evaluation shows
that a runtime-assurance claim rests on three things together: correct
obligations, complete mediation of consequential actions, and trustworthy
attribution of the evidence that discharges their preconditions.

\paragraph{Future work.}
Four directions follow from the limitations above. Integrate external engines
into the closed-loop harness, replacing the remaining reference implementations:
\phiS\ has never been executed on a stream engine here, and the full timed \phiT\
is checked offline on MonPoly rather than inside the loop. Repeat the closed-loop
harness against an optimised attacker of the kind that produces STAC's rates, and
against a weaker planner. Generate scenarios from independently specified mission
models rather than from seven families of our own, so that coverage is tested
beyond the structures we anticipated. And generalise the guarded composition
beyond its Boolean arming semantics, to uncertain and nested guards and a formal
treatment of guard provenance: the protected-status pattern suggests that guard
structure can support richer operators than the guarded maximum used here. That
last one is the most interesting formal question this work has raised.

\bibliographystyle{splncs04}
\bibliography{references}   

\begin{thebibliography}{10}
\providecommand{\url}[1]{\texttt{#1}}
\providecommand{\urlprefix}{URL }
\providecommand{\doi}[1]{https://doi.org/#1}

\bibitem{strel2017}
Bartocci, E., Bortolussi, L., Loreti, M., Nenzi, L.: Monitoring mobile and
  spatially distributed cyber-physical systems. Formal Methods and Models for
  System Design (MEMOCODE)  (2017). \doi{10.1145/3127041.3127050}

\bibitem{verimon2021}
Basin, D., Dardinier, T., Hauser, N., Heimes, L., Krsti{\'c}, S., Raszyk, M.,
  Schneider, J., Traytel, D., Zingg, S.: {VeriMon}: A formally verified
  monitoring tool. In: Theoretical Aspects of Computing (ICTAC). LNCS, vol.
  13572, pp.~1--6. Springer (2022). \doi{10.1007/978-3-031-17715-6_1}

\bibitem{monpoly2012}
Basin, D., Harvan, M., Klaedtke, F., Zalinescu, E.: {MONPOLY}: Monitoring
  usage-control policies. In: Runtime Verification (RV 2011). LNCS, vol.~7186,
  pp. 360--364. Springer (2012). \doi{10.1007/978-3-642-29860-8_27}

\bibitem{blsemantics2011}
Bauer, A., Leucker, M., Schallhart, C.: Runtime verification for {LTL} and
  {TLTL}. ACM Transactions on Software Engineering and Methodology
  \textbf{20}(4),  14:1--14:64 (2011). \doi{10.1145/2000799.2000800}

\bibitem{tessla2018}
Convent, L., Hungerecker, S., Leucker, M., et~al.: {TeSSLa}: Temporal
  stream-based specification language. In: Brazilian Symposium on Formal
  Methods (SBMF) (2018). \doi{10.1007/978-3-030-03044-5_10}

\bibitem{agentdojo2024}
Debenedetti, E., Zhang, J., Balunovi{\'c}, M., Beurer-Kellner, L., Fischer, M.,
  Tram{\`e}r, F.: {AgentDojo}: A dynamic environment to evaluate prompt
  injection attacks and defenses for {LLM} agents. In: NeurIPS Datasets and
  Benchmarks (2024)

\bibitem{thunderforge2025}
{Defense Innovation Unit}: Thunderforge: {AI} for {American} defense --- {AI}
  agents for military planning and operations. U.S. DoD prototype programme;
  initial mission partners U.S. Indo-Pacific and European Command (2025)

\bibitem{euaiact2024}
{European Parliament and Council}: Regulation (eu) 2024/1689 laying down
  harmonised rules on artificial intelligence (ai act). Official Journal of the
  European Union (2024), \url{http://data.europa.eu/eli/reg/2024/1689/oj}

\bibitem{rtlola}
Faymonville, P., Finkbeiner, B., Schledjewski, M., et~al.: {StreamLAB}:
  Stream-based monitoring of cyber-physical systems. In: Computer Aided
  Verification (CAV) (2019)

\bibitem{havelund2020timed}
Havelund, K., Peled, D.: First-order timed runtime verification using {BDD}s.
  In: Automated Technology for Verification and Analysis (ATVA). LNCS, vol.
  12302, pp. 3--24. Springer (2020). \doi{10.1007/978-3-030-59152-6_1}

\bibitem{havelund2018dejavu}
Havelund, K., Peled, D., Ulus, D.: First-order temporal logic monitoring with
  {BDD}s. In: Formal Methods in Computer Aided Design (FMCAD). pp. 116--123
  (2017). \doi{10.23919/FMCAD.2017.8102249}

\bibitem{llamaguard2023}
Inan, H., Upasani, K., Chi, J., et~al.: {Llama Guard}: {LLM}-based input-output
  safeguard for human-{AI} conversations. arXiv preprint arXiv:2312.06674
  (2023)

\bibitem{iso42001}
{International Organization for Standardization}: {ISO/IEC 42001:2023} ---
  information technology --- artificial intelligence --- management system.
  ISO/IEC 42001:2023 (2023), \url{https://www.iso.org/standard/81230.html}

\bibitem{tessla2024}
Kallwies, H., Leucker, M., Schmitz, M., et~al.: {TeSSLa}: An ecosystem for
  runtime verification. In: Runtime Verification (RV). LNCS, vol. 13498.
  Springer (2022). \doi{10.1007/978-3-031-17196-3_20}

\bibitem{kekatos2026csr}
Kekatos, N., Chintri, M., Katsaros, P., et~al.: Hierarchical security
  monitoring for edge-{IoT}: A formal methods approach. In: IEEE International
  Conference on Cyber Security and Resilience (CSR). IEEE (2026)

\bibitem{kekatos2026critis}
Kekatos, N., Chintri, M., Katsaros, P., et~al.: A resilient
  runtime-verification fabric for security monitoring of critical edge-{IoT}
  infrastructure. In: Critical Information Infrastructures Security (CRITIS).
  LNCS, Springer (2026)

\bibitem{kekatos2027ares}
Kekatos, N., Chintri, M., Katsaros, P., et~al.: Hybrid hierarchical runtime
  verification for edge-{IoT} security: Combining {MonPoly} and {RTLola}. In:
  Kieseberg, P., Skopik, F., Atli, B., Asplund, M. (eds.) Availability,
  Reliability and Security. ARES 2026 EU Projects Symposium Workshops. LNCS,
  vol. 16903, pp. 222--240. Springer, Cham (2027).
  \doi{10.1007/978-3-032-37218-5_14}

\bibitem{koutidis2026guardrails}
Koutidis, G., Kekatos, N., Nianios, T., Lekidis, A.: Constrained-action {AI}
  remediation for {SIEM}/{XDR} via a {NeMo-Guardrails} proxy. In: IEEE
  International Conference on Cyber Security and Resilience (CSR). IEEE (2026)

\bibitem{smartmedevac2024}
Krygier, J., Lubkowski, P., Maslanka, K., et~al.: Smart medical evacuation
  support system for the military. Sensors  \textbf{24}(14), ~4581 (2024).
  \doi{10.3390/s24144581}

\bibitem{stac2025}
Li, J.J., He, J., Shang, C., et~al.: {STAC}: When innocent tools form dangerous
  chains to jailbreak {LLM} agents (2025), arXiv:2509.25624

\bibitem{medevacdss2023}
Lubkowski, P., Krygier, J., Sondej, T., et~al.: Decision support system
  proposal for medical evacuations in military operations. Sensors
  \textbf{23}(11), ~5144 (2023). \doi{10.3390/s23115144}

\bibitem{sastl2020}
Ma, M., Bartocci, E., Lifland, E., Stankovic, J.A., Feng, L.: {SaSTL}: Spatial
  aggregation signal temporal logic for runtime monitoring in smart cities. In:
  Proceedings of the 11th ACM/IEEE International Conference on Cyber-Physical
  Systems (ICCPS) (2020). \doi{10.1109/ICCPS48487.2020.00013}

\bibitem{natoai2024}
{North Atlantic Treaty Organization}: Summary of the {NATO} revised artificial
  intelligence strategy.
  \url{https://www.nato.int/cps/en/natohq/official_texts_227237.htm} (2024),
  adopted 10 July 2024

\bibitem{conformal2024}
Quach, V., Fisch, A., Schuster, T., et~al.: Conformal language modeling. In:
  International Conference on Learning Representations (ICLR) (2024)

\bibitem{progent2025}
Shi, T., He, J., Wang, Z., et~al.: {Progent}: Programmable privilege control
  for {LLM} agents. arXiv preprint arXiv:2504.11703  (2025)

\bibitem{temperekidis2022fmirv}
Temperekidis, A., Kekatos, N., Katsaros, P.: Runtime verification for
  {FMI}-based co-simulation. In: Runtime Verification (RV). LNCS, vol. 13498.
  Springer (2022). \doi{10.1007/978-3-031-17196-3_19}

\bibitem{tsigkanos2022edge}
Tsigkanos, C., Bersani, M.M., Frangoudis, P.A., Dustdar, S.: Edge-based runtime
  verification for the {Internet of Things}. IEEE Trans. Serv. Comput.
  \textbf{15}(5),  2713--2727 (2022). \doi{10.1109/TSC.2021.3074956}

\bibitem{dodlima2024}
{U.S. Department of Defense, Chief Digital and Artificial Intelligence Office}:
  Task force lima executive summary. \url{https://www.ai.mil/} (2024)

\bibitem{agentspec2026}
Wang, H., Poskitt, C.M., Sun, J.: {AgentSpec}: Customizable runtime enforcement
  for safe and reliable {LLM} agents. In: Proceedings of the 48th IEEE/ACM
  International Conference on Software Engineering (ICSE) (2026)

\bibitem{guardagent2024}
Xiang, Z., Zheng, L., Li, Y., et~al.: {GuardAgent}: Safeguard {LLM} agents by a
  guard agent via knowledge-enabled reasoning. arXiv preprint arXiv:2406.09187
  (2024)

\bibitem{yin2024safeagentbench}
Yin, S., Pang, X., Ding, Y., et~al.: {SafeAgentBench}: A benchmark for safe
  task planning of embodied {LLM} agents. arXiv preprint arXiv:2412.13178
  (2024)

\end{thebibliography}

\end{document}